\documentclass[a4paper,12pt]{article}
\usepackage{amsmath,amsfonts,amsthm}

\def\Prob{{\Pr}}
\begin{document}
\title{The Three Doors Problem...\textbf{-s}\footnote{~ v.2 // arXiv.org:1002.3878 [stat.AP] // Submitted to \emph{Springer Lexicon of Statistics}}}
\author{Richard D. Gill\\ {\small Mathematical Institute, University Leiden, Netherlands}\\ {\small \texttt{http://www.math.leidenuniv.nl/$\sim$gill}}}
\date{March 1, 2010}

\maketitle

\raggedbottom

\section{Introduction}
\noindent The \emph{Three Doors Problem}, or \emph{Monty Hall Problem}, is familiar to statisticians as a paradox in elementary probability theory often found in elementary probability texts (especially in their exercises sections). In that context it is usually meant to be solved by careful (and elementary) application of Bayes' theorem. However, in different forms, it is much discussed and argued about and written about by psychologists, game-theorists and mathematical economists, educationalists, journalists, lay persons, blog-writers, wikipedia editors.  

In this article I will briefly survey the history of the problem and some of the approaches to it which have been proposed. My take-home message to you, dear reader,  is that one should distinguish two levels to the problem. 

There is an informally stated problem which you could pose to a friend at a party; and there are many concrete \emph{versions} or \emph{realizations} of the problem, which are actually the result of mathematical or probabilistic or statistical \emph{modelling}. This modelling often involves adding supplementary assumptions chosen to make the problem well posed in the terms of the modeller. The modeller finds those assumptions perfectly natural. His or her students are supposed to guess those assumptions from various key words (like:  ``indistinguishable'', ``unknown'') strategically placed in the problem re-statement. Teaching statistics is often about teaching the students to read the teacher's mind. Mathematical (probabilistic, statistical) modelling is, unfortunately, often solution driven rather than problem driven. 

The very same criticism can, and should, be levelled at this very article! By cunningly presenting the history of \emph{The Three Doors Problem} from my rather special point of view, I have engineered complex reality so as to convert the \emph{Three Doors Problem} into an illustration of my personal Philosophy of Science, my Philosophy of Statistics.

This means that I  have re-engineered the \emph{Three Doors Problem} into an example of the point of view that Applied Statisticians should always be wary of the lure of \emph{Solution-driven Science}. Applied Statisticians are trained to know Applied Statistics, and are trained to know how to convert real world problems into statistics problems. That is fine. But the best Applied Statisticians know that Applied Statistics is not the only game in town. Applied Statisticians are merely some particular kind of Scientists. They know lots about modelling uncertainty, and about learning from more or less random data, but probably not much about anything else. The Real Scientist knows that there is not a universal \emph{disciplinary} approach to every problem. The \emph{Real Statistical Scientist} modestly and  persuasively and realistically offers what his or her discipline has to offer in synergy with others. 

To summarize, we must distinguish between:  
\begin{itemize}
\item[(0)] the \emph{Three-Doors-Problem} \textbf{Problem} [sic], which is to make sense of some real world question of a real person.

\item[(1)] a large number of solutions to this \emph{meta}-problem, i.e., the many Three-Doors-Problem  \emph{Problems}, which are competing mathematizations of the meta-problem (0). 

\end{itemize}

\noindent Each of the solutions at level (1)  can well have a number of different solutions: nice ones and ugly ones; correct ones and incorrect ones. In this article, I will discuss three level (1) solutions, i.e., three different Monty Hall problems; and try to give three short correct and attractive solutions. 

Now read on. Be critical, use your intellect, don't believe anything on authority, and certainly not on mine.  Especially, don't forget the problem at meta-meta-level ($-1$), not listed above. 

 \emph{C'est la vie}. 

\section{Starting Point}
I shall start not with the historical roots of the problem, but with the question which made the Three Doors Problem famous, even reaching the front page of the \emph{New~York~Times}.

Marilyn vos Savant (a woman allegedly with the highest IQ in the world) posed the \emph{Three Door Problem} or \emph{Monty Hall Problem} in her ``Ask Marilyn'' column in \emph{Parade} magazine (September, 1990, p.~16), as posed to her by a correspondent, a Mr.~Craig Whitaker. It was, quoting vos Savant literally, the following:
\begin{quote}
\emph{Suppose you're on a game show, and you're given the choice of three doors: Behind one door is a car; behind the others, goats. You pick a door, say No.~1, and the host, who knows what's behind the doors, opens another door, say No. 3, which has a goat. He then says to you, ``Do you want to pick door No.~2?'' Is it to your advantage to switch your choice?}
\end{quote}
Apparently, the problem refers to a real American TV quiz-show, with a real presenter, called Monty Hall.

The literature on the Monty Hall Problem is enormous. At the end of this article I shall simply list two references which for me have been especially valuable: a paper by Jeff Rosenthal (2008) and a book by Jason Rosenhouse (2009). The latter has a huge reference list and discusses the pre- and post-history of vos Savant's problem. 

Briefly regarding the pre-history, one may trace the problem back through a 1975 letter to the editor in the journal \emph{The American Statistician} by biostatistician Steve Selkin,  to a problem called \emph{The Three Prisoners Problem} posed by Stephen Gardner in his Mathematical Games column in  \emph{Scientific American} in 1959, and from there back to \emph{Bertrand's Box Problem} in his 1889 text on Probability Theory. The internet encyclopedia \texttt{wikipedia.org} discussion pages (in many languages) are a fabulous though every-changing resource. Almost everything that I write here was learnt from those pages.

Despite making homage here to the two cited authors Rosenthal (2008) and Rosenhouse (2009) for their wonderful work, I emphasize that I strongly disagree with both Rosenhouse (``the canonical problem'') and Rosenthal (``the original problem'') on what the essential Monty Hall problem is.  I am more angry with certain other authors, who will remain nameless but for the sake of argument I'll just call Morgan \emph{et al.},\  for unilaterally declaring in \emph{The American Statistician} in 1981 \emph{their} Monty Hall problem to be the only possible sensible problem, for calling everyone who solved different problems stupid, and for getting an incorrect theorem\footnote{I refer to their result about the situation when we do not know the quiz-master's probability of opening a particular door when he has a choice, and put a uniform prior on this probability.}  published in the peer-reviewed literature.

Deciding unilaterally (Rosenhouse, 2009) that a certain formulation is \emph{canonical} is asking for a schism and for excommunication. Calling a particular version \emph{original} (Rosenthal, 2008) is asking for a historical contradiction. In view of the pre-history of the problem, the notion is not well defined. Monty Hall is part of folk-culture, culture is alive, the Monty Hall problem is not owned by a particular kind of mathematician who looks at such a problem from a particular point of view, and who adds for them ``natural'' extra assumptions which merely have the role of allowing their solution to work. Presenting any ``canonical'' or ``original'' Monty Hall problem together with a solution, is an example of \emph{solution driven science} --- you have learnt a clever trick and want to show that it solves lots of problems.

\section{Three Monty Hall Problems}

I will concentrate on three different particular Monty Hall problems. One of them (Q-0)  is simply to answer the question literally posed by Marilyn vos Savant, ``would you switch?''. The other two (Q-1, Q-2) are popular mathematizations, particularly popular among experts or teachers of elementary probability theory: one asks for the unconditional probability that ``always switching'' would gets the car, the other asks for the conditional probability given the choices made so far. Here they are:
\begin{itemize}
\item[Q-0:] Marilyn vos Savant's (or Craig Whitaker's) question ``\emph{Is it to your advantage to switch?}''
\item[Q-1:] A mathematician's question``\emph{What is the unconditional probability that switching gives the car?}''
\item[Q-2:] A mathematician's question ``\emph{What is the conditional probability that switching gives the car, given everything so far?}''
\end{itemize}
The free, and freely editable, internet encyclopedia Wikipedia is the scene of a furious debate as to which mathematization Q-1 or Q-2 is the right starting point for answering the verbal question Q-0 (to be honest, many of the actors claim another ``original'' question as \emph{the} original question). Alongside that, there is a furious debate as to which  supplementary conditions are obviously implicitly being made. For each protagonist in the debate, those are the assumptions which ensure that his or her question has a unique and nice answer. My own humble opinion is ``neither Q-1 nor Q-2, though the unconditional approach comes closer'. I prefer Q-0,  and I prefer to see it as a question of \emph{game theory} for which, to my mind, [almost] no supplementary conditions need to be made. 

Here I admit that I will suppose that the player knows game-theory and came to the quiz-show prepared. I will also suppose that the player wants to get the Cadillac while Monty Hall, the quizmaster, wants to keep it.

My analysis below of both problems Q-1 and Q-2 yields the good answer ``$2/3$'' under minimal assumptions, and almost without computation or algebraic manipulation. I will use Israeli (formerly Soviet Union) mathematician Boris Tsirelson's proposal on Wikipedia talk pages to use symmetry to deduce the conditional probability from the unconditional one. (Boris graciously gave me permission to cite him here, but this should not be interpreted to mean that anything written here also has his approval). 

You, the reader, may well prefer a calculation using Bayes' theorem, or a calculation using the definition of conditional probability; I think this is a matter of taste.

I finally use a game-theoretic point of view, and von Neumann's minimax theorem, to answer the question Q-0 posed by Marilyn vos Savant, on the assumptions just stated.

Let the three doors be numbered in advance $1$, $2$, and $3$. I add the universally agreed (and historically correct)additional assumptions: Monty Hall knows in advance where the car is hidden, Monty Hall always opens a door revealing a goat. 

Introduce four random variables taking values in the set of door-numbers $\{1,2,3\}$:
\begin{itemize}
\item[]$C$: the quiz-team hides the Car (a Cadillac) behind door $C$,
\item[]$P$: the Player chooses door $P$,
\item[]$Q$: the Quizmaster (Monty Hall) opens door $Q$,
\item[]$S$: Monty Hall asks the player if she'ld like to Switch to door $S$.
\end{itemize}
Because of the standard story of the Monty Hall show, we certainly have:
\begin{itemize}
\item[] $Q \ne P$, the quizmaster \emph{always} opens a door different to the player's first choice,
\item[] $Q \ne C$, opening that door \emph{always} reveals a goat,
\item[] $S \ne P$, the player is \emph{always} invited to switch to another door,
\item[] $S \ne Q$, \emph{no} player wants to go home with a goat.
\end{itemize}
It does not matter for the subsequent mathematical analysis whether probabilities are subjective (Bayesian) or objective (frequentist); nor does it matter whose probabilities they are supposed to be, at what stage of the game. Some writers think of the player's initial choice as fixed. For them, $P$ is degenerate. 

I simply merely down some mathematical assumptions and deduce mathematical consequences of them. 

\section{Solution to Q-1:\\ unconditional chance that switching wins}

By the rules of the game and the definition of $S$, if $P\ne C$ then $S =C$, and vice-versa. A ``switcher'' would win the car if and only if a ``stayer'' would lose it. Therefore:
\begin{quotation}
\noindent \emph{If $\Prob(P=C)=1/3$ then $\Prob(S=C)=2/3$, since the two events are complementary.}
\end{quotation}

\section{Solution to Q-2:\\ probability car is behind door 2 given \\ you chose door 1, Monty Hall opened 3}

First of all, suppose that $P$ and $C$ are uniform and independent, and that given $(P,C)$, suppose that $Q$ is uniform on its possible values (unequal to those of $P$ and of $C$). Let $S$ be defined as before, as the third door-number different from $P$ and $Q$.  The joint law of $C,P,Q,S$ is by this definition invariant under renumberings of the three doors. Hence $\Prob(S=C|P=x,Q=y)$ is the same for all $x\ne y$. By the law of total probability,  $\Prob(S=C)$ (which is equal to $2/3$ by our solution to Q-1) is equal to the weighted average of all  $\Prob(S=C|P=x,Q=y)$, $x\ne y \in\{1,2,,3\}$.  Since the latter are all equal, all these six conditional probabilities are equal to their average $2/3$.

Conditioning on $P=x$, say, and letting $y$ and $y'$ denote the remaining two door numbers, we find the following corollary:

Now take the door chosen by the player as fixed, $P\equiv 1$, say. We are to compute $\Prob(S=C|Q=3)$. \emph{Assume that all doors are equally likely to hide the car and assume that the quizmaster chooses completely at random when he has a choice}. Without loss of generality we may as well pretend that $P$ was chosen in advance completely at random. Now we have embedded our problem into the situation just solved, where $P$ and $C$ are uniform and independent.

\begin{quotation}
\noindent \emph{If $P\equiv 1$ is fixed, $C$ is uniform, and $Q$ is symmetric, then ``switching gives car'' is independent of quizmaster's choice, hence}
\end{quotation}
$$\Prob(S=C|Q=3)~=~\Prob(S=C|Q=2')~=~\Prob(S=C)~=~2/3.$$

Some readers may prefer a direct calculation. Using Bayes' theorem in the form ``posterior odds equal prior odds times likelihoods'' is a particularly efficient way to do this. The probabilities and conditional probabilities below are all conditional on  $P=1$, or if your prefer with $P\equiv 1$. 

We have uniform prior odds $$\Prob(C=1):\Prob(C=2):\Prob(C=3)~=~1:1:1.$$
The likelihood for $C$, the location of the car, given data $Q=3$, is (proportional to) the discrete density function of $Q$ given $C$ (and $P$)
$$\Prob(Q=3|C=1):\Prob(Q=3|C=2):\Prob(Q=3|C=3)~=~\frac 1 2:1:0.$$
The posterior odds are therefore proportional to the likelihood. It follows that the posterior probabilities are
$$\Prob(Q=3|C=1)=\frac 13,\quad \Prob(Q=3|C=2)=\frac 2 3,\quad \Prob(Q=3|C=3)=0.$$

\section{Answer to Marylin vos Savant's Q-0: \\
should \emph{you} switch doors?}

\noindent Yes. Recall, \emph{You only know that Monty Hall always opens a door revealing a goat}.   You didn't know what strategy the quiz-team and quizmaster were going to use for their choices of the distribution of $C$ and the distribution of $Q$ given $P$ and $C$, so naturally (since you know elementary Game Theory) you had picked your door uniformly at random. Your strategy of choosing $C$ uniformly at random guarantees that $\Prob(C=P)=1/3$ and hence that $\Prob(S=C)=2/3$. 

It was easy for you to find out that this combined strategy, which I'll call ``symmetrize and switch'',  is your so-called minimax strategy. 

On the one hand, ``symmetrize and switch''  guarantees you a $2/3$ (unconditional) chance of winning the car, whatever strategy used by the quizmaster and his team.

On the other hand,  if the quizmaster and his team use their ``symmetric'' strategy ``hide the car uniformly at random and toss a fair coin to open a door if there is choice'', then you cannot win the car with a \emph{better} probability than $2/3$. 

The fact that your ``symmetrize and switch'' strategy gives you ``at least'' $2/3$, while the quizmaster's ``symmetry'' strategy prevents you from doing better, proves that these are the respective minimax strategies, and $2/3$ is the game-theoretic value of this two-party zero-sum game. (Minimax strategies and the accompanying ``value'' of the game exist by virtue of John von Neumann's (1929) minimax theorem for finite two-party zero-sum games).

There is not much point for you in worrying about your conditional probability of winning conditional on your specific initial choice and the specific door opened by the quizmaster, say doors $1$ and $3$ respectively. You don't know this conditional probability anyway, since you don't know the strategy used by quiz-team and the quizmaster. (Even though you know probability theory and game theory, they maybe don't). However, it is maybe comforting to learn, by easy calculation, that if the car is hidden uniformly at random, then your conditional probability cannot be \emph{smaller} than $1/2$. So in that case at least, it certainly never \emph{hurts} to switch door.

\section{Discussion} 
Above I tried to give short clear mathematical solutions to three mathematical problems. Two of them were problems of elementary probability theory, the third is a problem of elementary game theory. As such, it involves not much more than elementary probability theory and the beautiful minimax theorem of John von Neumann (1928). That a finite two-party zero-sum game has a saddle-point, or in other words, that the two parties in such a game have matching minimax strategies (if randomization is allowed), is not obvious. It seems to me that probabilists ought to know more about game theory, since every ordinary non-mathematician who hears about the problem starts to wonder whether the quiz-master is trying to cheat the player, leading to an infinite regress: if I know that he knows that I know that.... 

I am told that the literature of mathematical economics and of game theory is full of Monty Hall examples, but no-one can give me a nice reference to a nice game-theoretic solution of the problem. Probably game-theorists like to keep their clever ideas to themselves, so as to make money from playing the game. Only losers write books explaining how the reader could make money from game theory.

It would certainly be interesting to investigate more complex game-theoretic versions of the problem. If we take Monty Hall as a separate player to the TV station, and note that TV ratings are probably helped if nice players win while annoying players lose, we leave elementary game theory and must learn the theory of Nash equilibria.

Then there is a sociological or historical question: who ``owns'' the Monty Hall problem? I think the answer is obvious: no-one. A beautiful mathematical paradox, once launched into the real world, lives it own life, it evolves, it is re-evaluated by generation after generation. This point of view actually makes me believe that Question 0: \emph{would you switch} is the right question, and no further information should be given beyond the fact that you know that the quizmaster knows where the car is hidden, and always opens a door exhibiting a goat. Question 0 is a question you can ask a non-mathematician at a party, and if they have not heard of the problem before, they'll give the wrong answer (or rather, one of the two wrong answers: \emph{no} because nothing is changed, or \emph{it doesn't matter} because it's now 50--50). My mother, who was one of Turing's computers at Bletchley Park during the war, but who had had almost no schooling and in particular never learnt any mathematics, is the only person I know who immediately said: \emph{switch}, by immediate intuitive consideration of the 100-door variant of the problem. The problem is a \emph{paradox} since you can next immediately convince anyone (except lawyers, as was shown by an experiment in Nijmegen), that their initial answer is wrong.

The mathematizations Questions 1 and 2 are not (in my humble opinion!) \emph{the} Monty Hall problem; they are questions which probabilists might ask, anxious to show off Bayes' theorem or whatever. Some people intuitively try to answer Question 0 via Questions 1 and 2; that is natural, I do admit. And sometimes people become very confused when they realize that the answer to Question 2 can only be given its pretty answer ``2/3'' under further conditions. It is interesting how in the pedagogical mathematical literature, the further conditions are as it were held under your nose, e.g. by saying ``three \emph{identical} doors'',  or replacing Marilyn's "say, door 1'' by the more emphatic ``door 1''.

It seems to me that adding into the question explicitly the remarks that the three doors are equally likely to hide the car, and that when the quizmaster has a choice he secretly tosses a fair coin to decide, convert this beautiful paradox into a probability puzzle with little appeal any more to non experts.

It also converts the problem into one version of the three prisoner's paradox. The three prisoners problem is isomorphic to the conditional probabilistic three doors problem.  I always found it a bit silly and not very interesting, but possibly that problem too should be approached from a sophisticated game theoretic point of view.

By the way, Marilyn vos Savant's original question is semantically ambiguous, though this might not be noticed by a non-native English speaker. Are the mentioned door numbers, huge painted numbers on the front of the doors \emph{a priori}, or are we just for convenience \emph{naming} the doors by the choices of the actors in our game \emph{a posteriori}. Marilyn stated in a later column in \emph{Parade} that she had originally been thinking of the latter. However, her own offered solutions are not consistent with a single unambiguous formulation. Probably she did not find the difference very interesting.

This little article contains nothing new, and only almost trivial mathematics. It is a plea for future generations to preserve the life of \emph{The True Monty Hall paradox}, and not let themselves be misled by probability purists who say ``you \emph{must} compute a conditional probability''.

\newpage

\section*{References} ~
\raggedright 
\frenchspacing 
\parskip 0.2 cm
\leftskip 0.5 cm
\parindent -0.5 cm

Gill, Richard D. (2010), The one and only true Monty Hall probem, submitted to \emph{Statistica Neerlandica}. \texttt{arXiv.org:1002.0651 [math.HO]}

Rosenhouse, Jason (2009), \emph{The Monty Hall Problem}, Oxford University Press.

Rosenthal, Jeffrey S. (2008), Monty Hall, Monty Fall, Monty Crawl, \emph{Math Horizons}, September 2008, 5--7. Reprint: \texttt{http://probability.ca/jeff/writing/montyfall.pdf}

\end{document}